# Thermal and electrical conductivity of iron at Earth's core conditions

Monica Pozzo[a], Chris Davies[c], David Gubbins[c,d] and Dario Alfè[a,b]

[a]Department of Earth Sciences and Thomas Young Centre @ UCL, UCL, Gower Street, WC1E 6BT, London, U.K.
[b]Department of Physics and Astronomy and London Centre for Nanotechnology, UCL, Gower Street, WC1E 6BT, London, U.K.
[c]School of Earth & Environment, University of Leeds, Leeds LS2 9JT, UK
[d]Institute of Geophysics and Planetary Physics, Scripps Institution of Oceanography
University of California at San Diego, 9500 Gilman Drive #0225, La Jolla, CA 92093-0225

**The Earth acts as a gigantic heat engine driven by decay of radiogenic isotopes and slow cooling, which gives rise to plate tectonics, volcanoes, and mountain building. Another key product is the geomagnetic field, generated in the liquid iron core by a dynamo running on heat released by cooling and freezing to grow the solid inner core, and on chemical convection due to light elements expelled from the liquid on freezing. The power supplied to the geodynamo, measured by the heat-flux across the core-mantle boundary (CMB), places constraints on Earth's evolution[1]. Estimates of CMB heat-flux[2-5] depend on properties of iron mixtures under the extreme pressure and temperature conditions in the core, most critically on the thermal and electrical conductivities. These quantities remain poorly known because of inherent difficulties in experimentation and theory. Here we use density functional theory to compute these conductivities in liquid iron mixtures at core conditions from first principles- the first directly computed values that do not rely on estimates based on extrapolations. The mixtures of Fe, O, S, and Si are taken from earlier work[6] and fit the seismologically-determined core density and inner-core boundary density jump[7,8]. We find both conductivities to be 2-3 times higher than estimates in current use. The changes are so large that core thermal histories and power requirements must be reassessed. New estimates of adiabatic heat-flux give 15-16 TW at the CMB, higher than present estimates of CMB heat-flux based on mantle convection[1]; the top of the core must be thermally stratified and any convection in the upper core driven by chemical convection against the adverse thermal buoyancy or lateral variations in CMB heat flow. Power for the geodynamo is greatly restricted and future models of mantle evolution must incorporate a high CMB heat-flux and explain recent formation of the inner core.**

First principles calculations of transport properties based on density functional theory (DFT) have been used in the past for a number of materials (e.g. [9,10]). Recently, increased computer power has facilitated simulations of large systems, allowing to completely address the problem of the size of the simulation cell, which for the electrical conductivity ($\sigma$) can be a serious one[11].
Here we report a series of calculations of the electrical and thermal conductivity ($k$) of iron at Earth's core conditions, using DFT. We previously used these methods to compute an extensive number of thermodynamic properties of iron and iron alloys, including the whole melting curve of iron in the pressure range [50-400] GPa[12,13] and the chemical potentials of oxygen, sulphur and silicon in solid and liquid iron at inner core boundary (ICB) conditions, which we used to place constraints on core composition[6]. Recently, we computed the conductivity of iron at ambient conditions, and obtained values in very good agreement with experiments[14].

The calculations of the conductivities were performed at 7 points on the iron and two possible core adiabats. These are determined by taking the temperature at the ICB to be 6350 K (melting temperature of pure iron)[13], 5700 K (melting temperature of the mixture with 10% Si and 8 % O, corresponding to an inner-core density jump $\Delta\rho = 0.6$ *gm cc$^{-1}$*)[6] and 5500 K (melting temperature of the mixture with 8% Si and 13% O, corresponding to $\Delta\rho = 0.8$ *gm cc$^{-1}$*)[6] for the three cases respectively, and following the line of constant entropy as the pressure is reduced to that of the CMB. Results are reported in Fig 1, and show a smooth variation of these parameters in the core; $\sigma$ only varies by ~ 13 % between the ICB and the CMB, and it is almost the same for all adiabats. A recent shock wave (SW) experiment[15] reported $\sigma = 0.765 \times 10^6$ $\Omega^{-1}$m$^{-1}$ for pure iron at 208 GPa, and an older SW measurement[16] reported $\sigma = 1.48 \times 10^6$ $\Omega^{-1}$m$^{-1}$ at 140 GPa. Our values are closer to the latter. *k* has a larger variation, as implied by the Wiedemann-Franz law, which we found to be closely followed throughout the core with a Lorenz parameter L = 2.48 – 2.51 x $10^{-8}$ W Ω $K^{-2}$. The ionic contribution to *k* was calculated using the classical potential used as a reference system in Ref. [12], which was shown to describe very accurately the energetics of the system and the structural and dynamical properties of liquid iron at Earth's core conditions. We found that the ionic contribution is only between 2.5 and 4 W $m^{-1}$ $K^{-1}$ on the adiabat, which is negligible compared to the electronic contribution, as expected.

The *k* estimates in Fig 1 are substantially larger than previously used in the geophysical literature, approximately doubling the heat conducted down the adiabatic gradient in the core and halving the power to drive a dynamo generating the same magnetic field. These considerations demand a revision of the power requirements for the geodynamo. The conductivities for liquid mixtures appropriate to the outer core are likely to be smaller than for pure iron, preliminary calculations suggesting about 30% lower, a smaller difference than that found in previous work[17], but in close agreement with extrapolations obtained from recent DAC experiments, which reported a value in the range 90-130 W $m^{-1}K^{-1}$ at the top of the outer core[18]. Our values are also in broad agreement with recently reported DFT calculations[19].

We focus on estimates for the two mixtures above, corresponding to ICB density jumps 0.6 gm/cc[8] and 0.8 gm/cc[7]. There is relatively little effect on the conductivities in the two cases, because any additional O in the outer core must be balanced by less S or Si to maintain the mass of the whole core, which is well constrained. The larger density jump gives a higher O content, more gravitational energy, a lower ICB temperature and lower adiabatic gradient: it therefore favours compositional over thermal convection. The relevant values are given in Table 1.

We estimate power requirements for the dynamo using the model described in a previous study [Ref. 5, and Online Methods]. Neglecting small terms, the total CMB heat-flux, $Q_{CMB}$, is the sum of proportional to either the CMB cooling rate, $dT_0/dt$, or the amount of radiogenic heating, $h$: $Q_{CMB} = Q_s + Q_L + Q_g + Q_r$, where the terms on the right-hand side represent respectively the effects of secular cooling, latent heat, gravitational energy and radiogenic heating. The cooling rate, expressed in degrees per billion years, can be varied with the radiogenic heating to produce some desired outcome: a fixed mantle heat-flux, a marginal dynamo (no entropy left for Ohmic dissipation, $E_\sigma$), or a primordial inner core (by decreasing the cooling rate and increasing the radiogenic heating). Results for a suite of 11 models are shown in Table 2.

Model 1 fails as a dynamo, there is an entropy deficit, meaning the assumption that the whole core can convect is incorrect – the temperature gradient must fall below the adiabat to balance the entropy equation. A dynamo might still be possible with a large part of the core completely stratified. Model 2 demonstrates the efficiency of compositional convection: the entropy is greatly increased compared to Model 1 with no change in cooling rate and little increase in heat-flux; the dynamo is now marginal. Model 3 has an increased cooling rate and consequent younger inner core to demonstrate what is required for a marginal dynamo with $\Delta\rho = 0.6$. Models 4 and 5 have cooling rates that make the CMB thermally neutral; the CMB heat-flux is equal to that conducted down the adiabat. Models 6 and 7 have some radiogenic heating and the original cooling rate and operate as dynamos, although they are still thermally stable at the top of the core. Models 8-11 have cooling rates that yield old inner core ages, 3.5 and 4.5 Ga, and the radiogenic heating has been adjusted to make a marginal dynamo. They are also thermally stable at the top of the core.

We estimate stable layer thicknesses by computing the radial variation of thermal and compositional gradients for each model using the equations of a previous study[Ref. 20, Online Methods], which are derived from the equations of core energetics[5]. To compare thermal and chemical gradients we multiply the latter by the ratio of compositional and thermal expansion coefficients $\alpha_c/\alpha_T$, thereby converting compositional effects into equivalent thermal effects. The base of the stable layer is taken defined as the point where the stabilising adiabatic gradient, $T'_a$, crosses the combined destabilising gradient $T' = T'_L + T'_s + T'_c + T'_r$, where the terms represent respectively latent heat, secular cooling, compositional buoyancy and radiogenic heating.

Stable layer thicknesses are hundreds of km in all models except those with cooling rates that are so rapid as to make the inner core too young; without compositional buoyancy the layers in all models except 4 and 5 span half the core (Table 2). Radiogenic heating thins the layers for the same cooling rate. Profiles of stabilising and destabilising gradients (Fig. 2) show that destabilising gradients are greatest at depth, but much reduced compared to previous models[20] because they each depend on a factor $1/k$. The thermal conductivity increases by 50% across the core, increasing the heat conducted down the adiabat at depth, further reducing the power available to drive convection at depth. Combined thermochemical profiles suggest that compositional buoyancy near the top of core is not strong enough to drive convection against the adverse temperature gradient.

Stable layers could be thinned or partially disturbed by convection, through penetration or instability, or some other effect not included in our simple model. A potentially more effective mechanism for inducing vertical mixing near the CMB is through lateral variations in CMB heat-flux, which can drive motions without having to overcome the gravitational force. The presence of lateral variations makes the relevant heat-flux for core mixing the maximum at the CMB[21], which could be as much as 10 times the average[22]; this does not influence dynamo entropy calculations but does allow magnetic flux to be carried to the surface in regions of cold mantle, as is observed[23].

As well as raising $k$, our calculations also raise $\sigma$ to about twice the current estimate. Two important quantities depend on $\sigma$: the magnetic diffusion time, or time taken for the slowest decaying dipole mode to fall by a factor $e$ in the absence of a dynamo, and the magnetic Reynolds number $Rm$, which measures the rate of generation of magnetic energy by a given flow. The magnetic diffusion time is increased to about 50 kyr. This may have significant implications for the theory of the secular variation: it makes the frozen flux approximation more accurate and lengthens the time scale of all diffusion-dominated processes, including polarity reversals. If current estimates of $Rm$ are appropriate for the core[24], the increased conductivity implies that the geodynamo can operate on slower fluid flows and less input power from thermal and compositional convection.

Revised estimates of $\sigma$ and $k$ calculated directly at core conditions for the first time, have fundamental consequences for the thermochemical evolution of the deep Earth. New estimates of the power requirements for the geodynamo suggest a CMB heat-flux in the upper range of what is considered reasonable for mantle convection unless very marginal dynamo action can be sustained, while a primordial inner core is only possible with a significant concentration of radiogenic elements in the core. There are objections to a high CMB heat-flux and also to radiogenic heating in the core[25-27], but one of the two seems inevitable if we are to have a dynamo. If the inner core is young these high values of conductivity provide further problems with maintaining a purely thermally-driven dynamo. A thermally stratified layer at the top of the core also appears inevitable. Viable thermal history models that produce thin stable layers and an inner core of ~1 Ga are likely to require a fairly rapid cooling rate and some radiogenic heating. The presence of a stable layer, and the effects associated with an increased electrical conductivity, have significant implications for our understanding of the geomagnetic secular variation.

**Methods summary**
Calculations were performed using DFT with the same technical parameters employed in Refs. [6,12-14]. We used the VASP code[28], PAW potentials[29] with $4s^13d^7$ valence configuration, the Perdew-Wang[30] functional, a plane wave cutoff of 293 eV, and single particle orbitals were occupied according to Fermi-Dirac statistics. We tested the inclusion in valence of semi-core $3s$ and $3p$ states on the conductivity and found that, as in the zero pressure case[14], it is completely negligible.
The electrical conductivity and the electrical component of the thermal conductivity have been calculated using the Kubo-Greenwood formula and the Chester-Thellung-Kubo-Greenwood formula as implemented in VASP[31].
Because of the low mass of the electrons compared to the ions, the conductivities may be calculated by assuming frozen ionic configurations, and averaging over a sufficiently large set representing the typical distribution of the ions at the pressures and temperatures of interest.

Molecular dynamics simulations were performed in the canonical ensemble using cubic simulation cells with 157 atoms and the Γ point, a time step of 1 fs, and an efficient extrapolation of the charge density which speeds up the simulations by roughly a factor of two[32]. Each state point was simulated for at least 6 ps, from which we discarded the first ps to allow for equilibration and used the last 5 ps to extract 40 configurations separated by 0.125 ps. This time interval is roughly two times longer than the correlation time, and therefore the configurations are statistically independent from each other. Because of the high temperatures involved, the conductivities converge quickly with respect to **k**-point sampling and size of the simulation cell[14], and we found that with a 157-atom cells and the single **k**-point (1/4,1/4/,1/4) the results are converged to better than 1%. The ionic component of the thermal conductivity was calculated using the Green-Kubo formula[33].

**Supplementary Information** is linked to the online version of the paper at www.nature.com/nature.

**Acknowledgements**
DG is supported by CSEDI grant EAR1065597 of the National Science Foundation. CD is supported by a Natural Environment Research Council personal fellowship, reference NE/H01571X/1. MP is supported by NERC grant NE/H02462X/1 to DA. Calculations have been performed on the U.K. national facility HECToR.

**Author Contribution** DA and DG designed the project. MP and DA performed the first principles calculations. CD and DG performed the thermal history and core stratification calculations. All authors discussed the results and commented on the manuscript.

**Author Information** Reprints and permissions information is available at www.nature.com/reprints. The authors declare that they have no competing financial interests. Correspondence and requests for materials should be addressed to Dario Alfè . (email: d.alfe@ucl.ac.uk).

**Tables**

| $\Delta\rho$ | $T_{ICB}$ | $T_{CMB}$ | $k_{ICB}$ | $k_{CMB}$ | $\sigma_{ICB}(\times 10^6)$ | $\sigma_{CMB}(\times 10^6)$ | %O | %S/Si |
|---|---|---|---|---|---|---|---|---|
| 0.6 | 5700 | 4186 | 150 (223) | 100 (144) | 1.25 (1.56) | 1.11(1.36) | 8 | 10 |
| 0.8 | 5500 | 4039 | 150 (215) | 100 (140) | 1.24(1.57) | 1.11(1.37) | 13 | 8 |

Table 1: Parameters used to estimate power requirements for the geodynamo. Values in parenthesis are for pure iron, other values are approximations for core mixtures. Units are gm cc$^{-1}$ for the ICB density jump $\Delta\rho$; K for the temperatures, $T$; W m$^{-1}$K$^{-1}$ for the thermal conductivity, $k$; $\Omega^{-1}$m$^{-1}$ for the electrical conductivity, $\sigma$; % for molar concentrations.

| Model | $\Delta\rho$ | $dT_0/dt$ | h | $Q_{ad}$ | $Q_{CMB}$ | IC age | $E_\sigma$ | $\Delta$ (km) |
|---|---|---|---|---|---|---|---|---|
| 1 | 0.6 | 46 | 0 | 15.7 | 5.8 | 0.9 | -111 | 1022 |
| 2 | 0.8 | 46 | 0 | 15.2 | 6.1 | 1.0 | 5 | 826 |
| 3 | 0.6 | 57 | 0 | 15.7 | 7.2 | 0.7 | -2 | 833 |
| 4 | 0.6 | 123 | 0 | 15.7 | 15.6 | 0.3 | 652 | 110 |
| 5 | 0.8 | 115 | 0 | 15.2 | 15.2 | 0.4 | 865 | 0 |
| 6 | 0.6 | 46 | 3.0 | 15.7 | 11.7 | 0.9 | 85 | 659 |
| 7 | 0.8 | 46 | 3.0 | 15.2 | 11.9 | 1.0 | 208 | 468 |
| 8 | 0.6 | 11.2 | 6.8 | 15.7 | 14.7 | 3.5 | -3 | 1257 |
| 9 | 0.6 | 8.7 | 6.9 | 15.7 | 14.5 | 4.5 | -1 | 1472 |
| 10 | 0.8 | 12.2 | 6.3 | 15.2 | 13.7 | 3.5 | 4 | 1000 |
| 11 | 0.8 | 9.5 | 6.6 | 15.2 | 14.1 | 4.5 | 2 | 1128 |

Table 2. Heat-flux and entropy for various models of cooling and radiogenic heating. $\Delta\rho$ is the density jump at the ICB in gm/cc; $dT_0/dt$ the cooling rate of the CMB in K Gyr$^{-1}$; $h$ the radiogenic heat source in pW kg$^{-1}$; $Q_{ad} = -4\pi k(dT_{ad}/dr)$ is the heat conducted down the adiabat in TW where $(dT_{ad}/dr)$ is the adiabatic gradient; $Q_{CMB}$ is the heat-flux across the CMB in TW; $E_\sigma$ is the entropy available for the dynamo and other diffusive processes in MW K$^{-1}$. Inner core age is shown in Ga; Stable layer thicknesses, $\Delta$, are given in kilometres below the CMB.

**Figures**

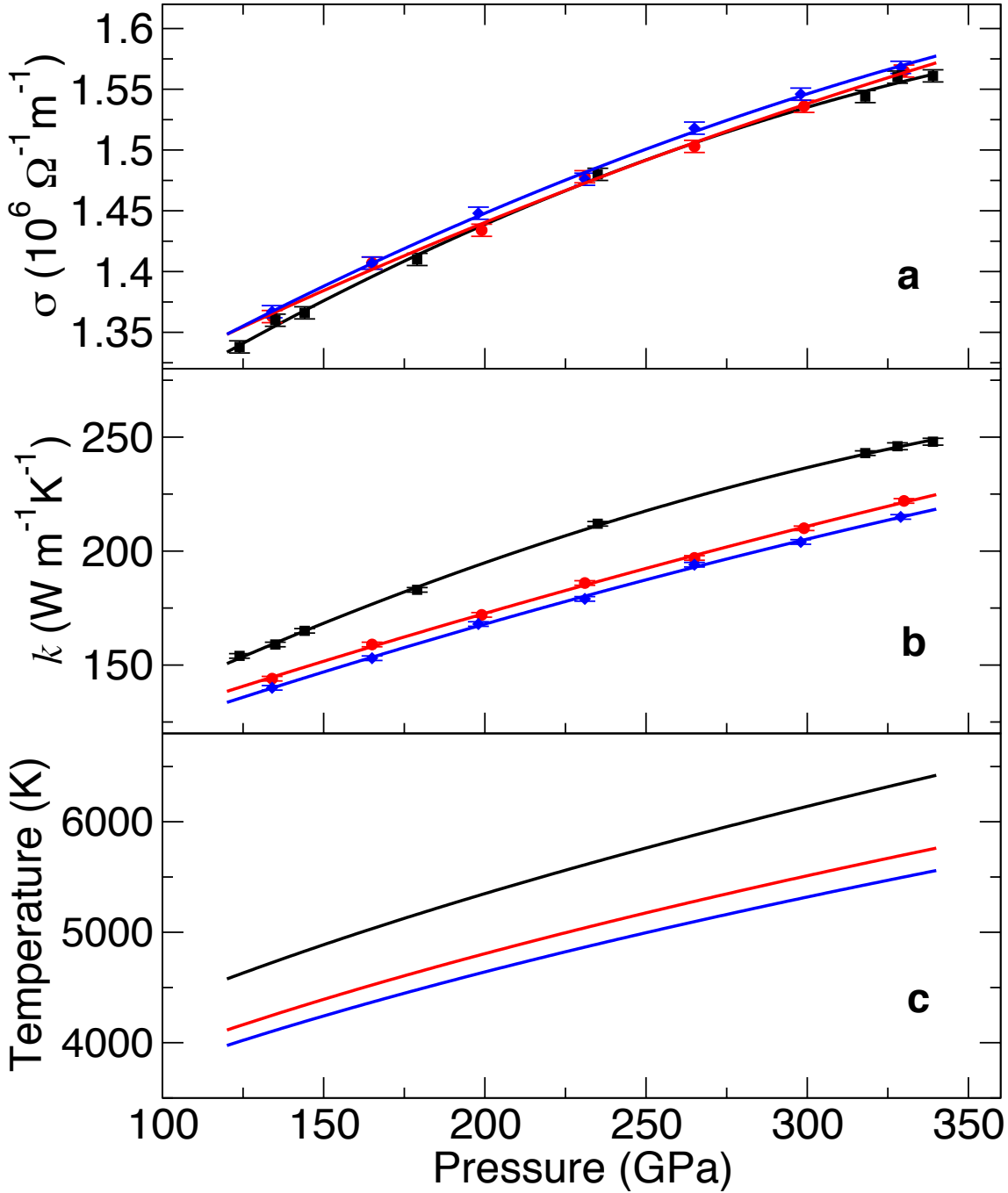

**Figure 1: Electrical and thermal conductivity of iron at Earth's outer core conditions.** Bottom panel (**c**): Adiabatic temperature profile in the outer core for three adiabats, corresponding to the melting temperature of pure iron at ICB pressure (black line), that of the mixture containing 10% Si and 8 % O (red line) and that of the mixture with 8% Si and 13 % O (blue line). Central panel (**b**) and top panel (**a**): electronic component of the thermal conductivity and electrical conductivity of pure iron, respectively, corresponding to the three adiabatic profiles displayed in bottom panel. Lines are quadratic fits to the first principles raw data (symbols). Error bars are estimated from the scattering of the data obtained from the 40 statistical independent configurations, and correspond to two standard deviations. Results are obtained with cells including 157 atoms and the single **k**-point (1/4,1/4,1/4), which are sufficient to obtain convergence within less than 1%.

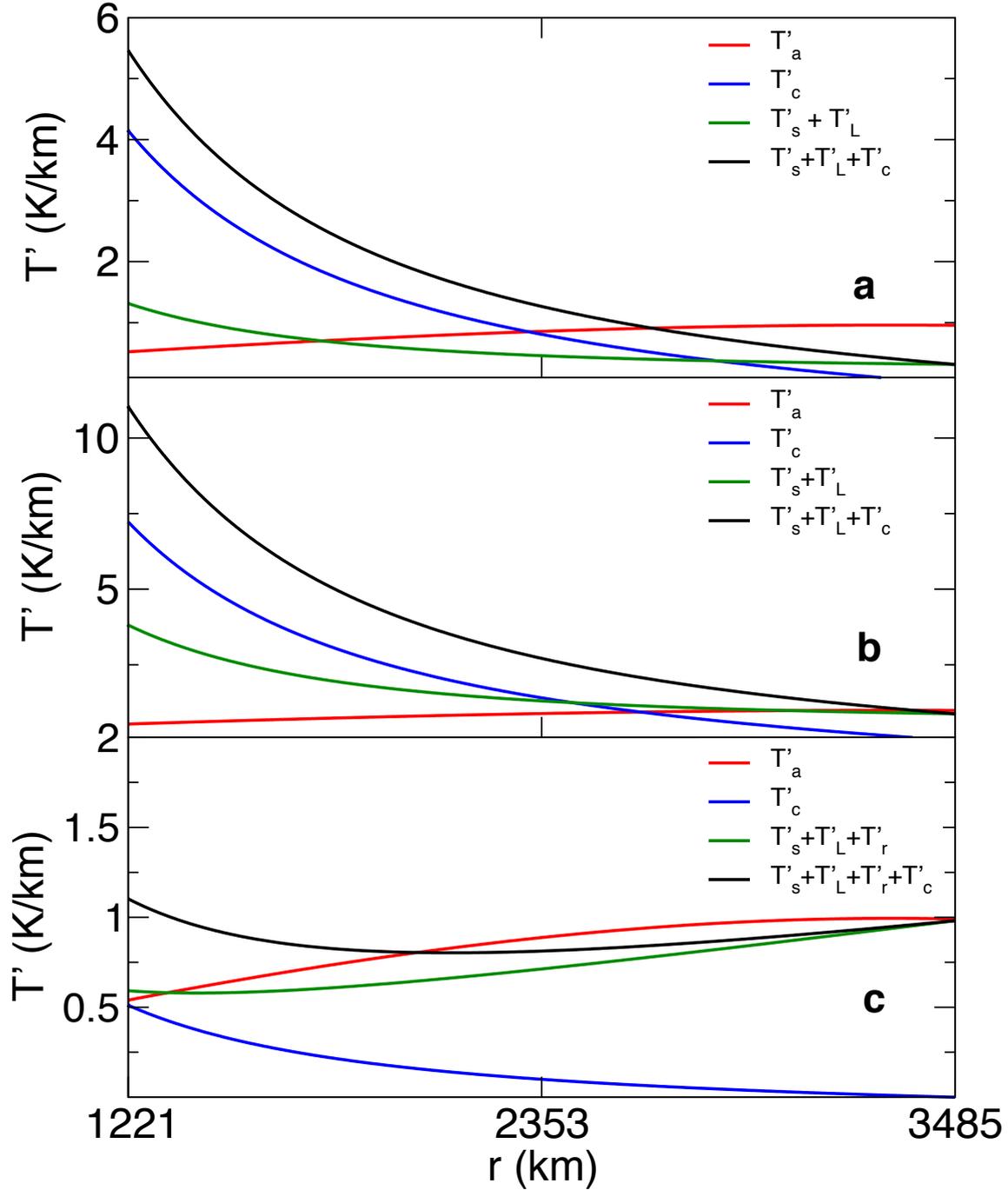

**Figure 2: Stabilising and destabilising gradients for 3 core energetics models.** Equivalent temperature gradients, T′, plotted against radius for three core evolution models. The stabilising gradient is due to conduction down the adiabat, $T'_a$ (red lines). Compositional buoyancy is denoted by $T'_c$ (blue lines), latent heat by $T'_L$, secular cooling by $T'_s$ and radiogenic heating by $T'_r$. The total destabilising thermal gradient is represented by the green lines; total destabilising thermochemical gradients are represented by black lines. Three models from Table 2 are shown: model 2 (**a**), $\Delta\rho$ = 0.8 gm cc$^{-1}$, $dT_o/dt$ = 46 K/Gyr and h = 0; model 4 (**b**), $\Delta\rho$ = 0.6 gm cc$^{-1}$, $dT_o/dt$ = 123 K/Gyr and h = 0; model 9 (**c**), $\Delta\rho$ = 0.6 gm cc$^{-1}$, $dT_o/dt$ = 8.7 K/Gyr and h = 6.9 pW/kg.

**Online Methods**

Power estimates for the geodynamo

Estimates of the power required to drive the geodynamo are obtained by considering the slow evolution of the Earth using equations describing the balances of energy and entropy in the core. A detailed derivation of these equations can be found in a previous study[5]. Conservation of energy simply equates the heat crossing the CMB to the sources within: specific heat of cooling $Q_s$, latent heat of freezing $Q_L$, radiogenic heating $Q_r$, gravitational energy loss $Q_g$, that is converted into heat by the frictional processes associated with the convection (almost entirely magnetic), and smaller terms[5] involving pressure changes and chemistry that we shall ignore:

$$Q_{CMB} = Q_s + Q_L + Q_g + Q_r. \qquad (1)$$

All terms on the right-hand side of (1) can be written in terms of either the cooling rate at the CMB, $dT_0/dt$, or the amount of radiogenic heating $h$. There is no dependence on the conductivities or the magnetic field, which are merely agents by which energy is converted to heat within the core.

These quantities do enter the entropy balance, however. This equation has dissipation terms from thermal and electrical conduction, plus viscosity and molecular diffusion. They are all positive because of the second law of thermodynamics. They are balanced by entropies associated with the power driving the convection: heat pumped in at a higher temperature and removed at a lower temperature ($T_{CMB}$) and gravitational energy that directly stirs the core and is converted to heat by frictional processes, the heat then being convected and conducted away. Note that entropy from heat is multiplied by a Carnot-like ``efficiency factor'' ($1/T_{out} - 1/T_{in}$) (latent heat is the most efficient because it is released at the highest temperature and removed at the lowest) while the gravitational energy is not, $E_g = Q_g/T_{ICB}$. Gravitational energy is more efficient at removing entropy and therefore more efficient than heat at generating magnetic field.

$$E = E_s + E_L + E_r + E_g = E_k + E_\sigma + E_\alpha, \qquad (2)$$

where the four terms on the left-hand side represent secular cooling, latent heat release, radiogenic heating and gravitational energy loss. Adiabatic conduction entropy, $E_k$, is easily estimated from the thermal conductivity and adiabatic gradient and is large, of order $10^8$ W/K. The new estimate of conductivity doubles older ones and the higher ICB temperatures increase it still further. Barodiffusion, $E_\alpha$, is the tendency for light elements to migrate down a pressure gradient and its associated entropy is significant but small, not exceeding 2.5 MW/K in any of our estimates. Diffusional processes associated with convection and the geodynamo also produce entropy, denoted $E_\sigma$, mainly in the small scales. This presents a problem in estimation because the dominant contribution comes from magnetic fields, fluid flows, temperature and compositional fluctuations that cannot be observed and, in many cases, cannot even be simulated numerically. A low value of the power required to drive the dynamo, 0.5 TW[34], was obtained from a numerical dynamo simulation[35], which at an average temperature of 5000 K translates into $E_\sigma = 10^7$ W/K, an order of magnitude lower than $E_k$, but the numerical simulation necessarily reduces small scale magnetic fields and the value for the Earth could be much larger. It may well be that future numerical simulations with higher resolution will have higher ohmic dissipation approaching $E_k$. Magnetic diffusivity is much larger than any other diffusivity in the core, by many orders of magnitude, and in numerical simulations the viscosity, thermal, and molecular diffusivities are replaced with turbulent values to account for unresolved, turbulent, small scale fields. Even so, the associated entropies remain much smaller than that associated with magnetic fields: they are generally ignored, although we should bear in mind that they are all positive and could make a contribution.

Parameter values used to calculate thermal contributions to the energy and entropy balances (1) and (2) are taken from Table 1 of a previous study[36], except for the thermal conductivity and the temperatures of the CMB and ICB, which are taken from the present study. Latent heat, $Q_L$, depends on $\tau$, the difference between the melting and adiabatic gradients at the ICB; the value for the former is taken to be 9 K/GPa[36], while the value of the latter is calculated from Figure 1 of this study. Parameter values used to calculate compositional terms differ slightly from previous work[5], owing to their use of different concentrations for the light elements O, Si and S in the outer core. Concentration enters the calculation of gravitational energy through equation (9) of Gubbins et al 2004[5] (equations from this paper will be denoted Gx), which, along with G8, is used to define $Q_g$ in equation G18. Note also $Q_g$ depends on $\tau$. The remaining changes affect the barodiffusion, $E_\alpha$, which makes a small contribution to the entropy budget (2); for completeness we list the new parameter values required to determine $E_\alpha$ in tables 1 and 2.

Estimating stable layer thicknesses

Radial profiles of the thermal and compositional energy sources that power the dynamo are determined using the equations of a previous study[20], which are derived from the energy balance appropriate for the outer core[5]. The radial profiles represent conductive solutions that satisfy the total CMB heat-flux boundary condition for the temperature, zero CMB mass-flux for the composition, and fixed temperature and composition at the ICB[20]. Superimposed on this basic state are the small fluctuations associated with core convection and the dynamo process.

These radial profiles apply to a Boussinesq fluid and hence neglect compressibility effects other than when they act to modify gravity. This necessitates the use of an approximate form for the adiabatic temperature, a simple choice being a quadratic equation expressed in terms of the ICB and CMB temperatures[19]. Despite these simplifications, the CMB heat-fluxes computed from equations (23)—(27) of the incompressible model[20] are in good agreement with those obtained from the original equations[5] (see Supplementary Table 3), while the quadratic approximation for the adiabat differs by at most 10 K from the full calculation shown in Figure 1.

Compositional buoyancy is at least as important for driving the geodynamo as thermal buoyancy[e.g 5] and so we require a means of comparing the two in radial profiles, which is readily achieved by multiplying the former by the ratio of compositional and thermal expansion coefficients, $\alpha_c/\alpha_T$. This simple device converts compositional effects into equivalent thermal effects, thereby allowing all sources of buoyancy to be combined; it is also related to the condition of neutral stability discussed below. (However, it must be understood that the compositional term resulting from this transformation has nothing to do with the gravitational energy, $Q_g$, which is neglected in the Boussinesq equations[37].) We use the common approach (e.g. [37]) of defining all fluxes that represent sources of buoyancy associated with the convection in terms of a turbulent diffusivity, which is assumed constant. By contrast, the heat-flux due to conduction down the adiabatic gradient and the equivalent thermal flux due to barodiffusion must be defined in terms of molecular quantities.

The depth variation of the molecular thermal conductivity obtained from the DFT results is readily incorporated into the formulation of previous work[20] (equations from this work will be denoted DGxx). We write $k = k(r)$ to express the radial variation of the molecular thermal conductivity; equation DG8 must then be replaced by $q_a = \nabla \cdot (k(r)\nabla T_a)$, where $\nabla T_a$ is calculated from equation DG12. $k(r)$ is well-approximated by a parabolic conductivity variation, $k(r) = ar^2 + br + c$, which we use to calculate the heat-flux down the adiabatic gradient.

To investigate the presence of a stable layer we use temperature gradients instead of heat-fluxes, which are calculated using equations (30)-(34) of a previous study[20] with $k(r)$ replacing $k$ in the numerator of equation (DG30). The parameter values are the same as those used to estimate power requirements above. We define the base of the stable layer to be the point of neutral stability as given by Schwarzchild's criterion[38]

$$\left(\frac{dT}{dr} - \frac{dT_a}{dr}\right) + \frac{\alpha_c}{\alpha_T}\left(\frac{dc}{dr}\right) = 0$$

where $dT/dr$ is the total temperature gradient, $dT_a/dr$ is the adiabatic temperature gradient and $dc/dr$ is the total compositional gradient. We write this condition as
$T' = T'_L + T'_s + T'_c + T'_r - T'_a = 0$, where the terms represent respectively latent heat, secular cooling, compositional buoyancy and radiogenic heating and prime indicates differentiation with respect to $r$ (the barodiffusive contribution to $dc/dr$ is very small and has been omitted). Possible deviations from the layer thicknesses we obtain using this definition can only be obtained by solving the complete dynamo equations with correct parameters for the Earth, which is impossible at present. We believe this to be the best definition of the base of the layer given the nature of our thermodynamic model.

**Supplementary tables**

|  | Equation | Units | O | Si |
|---|---|---|---|---|
| Atomic Weight | G01 |  | 16 | 28 |
| $\bar{c}$ | G01 | % | 0.08 | 0.1 |
| $C$ | G01 |  | 0.0256 | 0.056 |
| $(\partial\bar{\mu}/\partial\bar{c})_{P,T}$ | G36 | eV/atom | 5.32 | 4.26 |
| $\lambda$ | G31 | eV/atom | 3.25 | 3.5 |
| $(\partial\mu/\partial c)_{P,T}$ | G32 | J Kg$^{-1}$ | $16.2 \times 10^7$ | $4.77 \times 10^7$ |
| $\alpha_c$ | G18 |  | 1.1 | 0.87 |
| $D$ | G16 | m$^2$s | $1 \times 10^{-8}$ | $5 \times 10^{-9}$ |
| $\alpha_D$ | G16 | Kg m$^{-3}$ s | $0.70 \times 10^{-12}$ | $1.19 \times 10^{-12}$ |

**Supplementary Table 1:** Parameter values used to calculate the gravitational energy, $Q_g$, and barodiffusion, $E_\alpha$, for an inner-core density jump of $\Delta\rho = 0.6$ *gm cc$^{-1}$*. $c$ and $\bar{c}$ are respectively the mass and molar fractions of light element; $(\partial\bar{\mu}/\partial\bar{c})_{P,T}$ is the derivative of the chemical potential $\mu$ with respect to molar concentration at constant pressure $P$ and temperature $T$; $\lambda$ is the correction to the chemical potential derived from ideal solution theory; $(\partial\mu/\partial c)_{P,T}$ is the derivative of the chemical potential with respect to mass concentration; $\alpha_c$ is the compositional expansion coefficient; $D$ is the molecular diffusivity; $\alpha_D$ is a coefficient used to calculate $E_\alpha$. For this density jump, $\tau = 1.35 \times 10^{-4}$ K/m.

|  | Equation | Units | O | Si |
|---|---|---|---|---|
| Atomic Weight | G01 |  | 16 | 28 |
| $\bar{c}$ | G01 | % | 0.13 | 0.08 |
| $C$ | G01 |  | 0.0428 | 0.0461 |
| $(\partial\bar{\mu}/\partial\bar{c})_{P,T}$ | G36 | eV/atom | 3.16 | 5.14 |
| $\lambda$ | G31 | eV/atom | 3.25 | 3.5 |
| $(\partial\mu/\partial c)_{P,T}$ | G32 | J Kg$^{-1}$ | $11.7 \times 10^7$ | $5.16 \times 10^7$ |
| $\alpha_c$ | G18 |  | 1.1 | 0.87 |
| $D$ | G16 | m$^2$s | $1 \times 10^{-8}$ | $5 \times 10^{-9}$ |
| $\alpha_D$ | G16 | Kg m$^{-3}$ s | $0.97 \times 10^{-12}$ | $1.10 \times 10^{-12}$ |

**Supplementary Table 2:** Same as Supplementary Table 1 but for an inner-core density jump of $\Delta\rho = 0.8$ *gm cc$^{-1}$*. For this density jump, $\tau = 1.47 \times 10^{-4}$ K/m.

| Model | $Q_L$ | $Q_s$ | $Q_r$ | $Q_a$ |
|---|---|---|---|---|
| Davies and Gubbins, 2011[19] | 0.485 | 0.440 | 13.3 | 15.3 |
| Gubbins et al, 2004[5] | 0.490 | 0.449 | 13.4 | 15.7 |

**Supplementary Table 3**: CMB heat-fluxes using compressible[5] and incompressible[20] models for the outer core energy balance. Heat-fluxes are given in TW and computed for model 9 in Table 2 of the main text.